\documentclass[a4paper,11pt]{article}
\usepackage[utf8]{inputenc}
\usepackage{authblk}

\pdfoutput=1

\usepackage{times}
\usepackage{bm}
\usepackage{natbib}
\setcitestyle{authoryear}
\usepackage{tikz}
\usepackage{amsmath,amssymb}

\newtheorem{definition}{Definition}
\newtheorem{example}{Example}
\newtheorem{lemma}{Lemma}

\newtheorem{remark}{Remark}

\usepackage{geometry}
\usepackage{color,soul}

\begin{document}
\date{}
\title{Mixture Models: Building a Parameter Space} 
\author{Vahed Maroufy  \hspace{1cm}{\it and}\hspace{1cm}  Paul Marriott}
\affil{\small Department of Statistics and Actuarial Science, University of Waterloo}
\maketitle

\begin{abstract}
Despite the  flexibility and popularity of mixture models, their associated  parameter spaces  are  often difficult to represent
due to fundamental identification problems. This paper looks at a novel way of representing  such a   space for general mixtures
of exponential families,  where the  parameters  are  identifiable, interpretable,  and, due to a tractable  geometric  structure,
the space allows fast computational algorithms to be constructed. 
\end{abstract}


\section{Introduction}\label{Introduction}

 Mixtures of exponential family models
   have found application in almost all areas of statistics, see \cite{Lindsay1995}, \cite{everitt1996introduction}, \cite{Mclachlan2000} and \cite{Schlattmann2009}. At  their best  they  can achieve a   balance between parsimony,   flexibility and  interpretability --  the ideal of  parametric statistical  modelling.   Despite their ubiquity there are fundamental  open problems associated with  inference on such  models. Since the mixing mechanism is unobserved, a  very wide choice  of possibilities  is always available to the modeller: discrete and finite with known or unknown support,  discrete and infinite,  continuous,   or any plausible combination of these.  This gives rise to the first open problem; what is a good way to define a suitable  parameter space  in this class of models?  
  Other,  related,  problems  include  the difficulty of estimating the number of components,   possible unboundedness and  non-concavity of the   log-likelihood function,   non-finite  Fisher information, and boundary problems giving rise to non-standard analysis.   All these issues  are described in more detail  below.
This paper defines   a new solution to  first of these problems. We show how to construct a parameter space for general mixtures of exponential families, $\int f(x; \mu) dQ(\mu)$,  where the  parameters  are  identifiable, interpretable,  and, due to a tractable  geometric  structure,   the space allows fast computational algorithms to be constructed.

\subsection{Background}\label{Background}

Let $f(x; \mu)$ be a member of the exponential family. It will be convenient, but not essential to any of the results of
this paper, to parameterize with the mean parameter $\mu$. We will further assume that the dimension of $\mu$ is small
enough to  allow  underlying Laplace expansions to be reasonable, \cite{shun1995laplace}. A mixture over this family
would have the form $\int f(x; \mu) dQ(\mu)$ where $Q$ is the mixing distribution which, as stated above, can be very
general.  Since $Q$ may lie  in the set of all distributions the `parameter space' of this set of models is infinite
dimensional and very complex.  It is tempting to restrict $Q$ to be a finite discrete distribution indeed, as shown
by \cite{Lindsay1995}, the non-parametric maximum likelihood estimate of $Q$ lies in such a family.  Despite this,
as the following example clearly shows, this is too rich a class to be identified in a statistically meaningful way. 
\begin{example}\label{Norm_ex1} 
For this example let $f(x;\mu)= \phi(x; \mu,1)$, a normal distribution with unit variance. 
The QQ plot in Fig.~\ref{Norm_ex1_fig} compares two data sets generated from two different finite mixture models with five and ten components respectively. The plot shows that data generated from each  can have very similar empirical distributions --
thus it would be very hard to differentiate between these models and hence  estimate the number of components. In this example the components of the mixing distributions have been selected to be close to one another and to have the same lower order moment structure. 
\begin{figure}[!h]  
  \begin{center}
  \includegraphics[width=0.37\textwidth,natwidth=510,natheight=500]{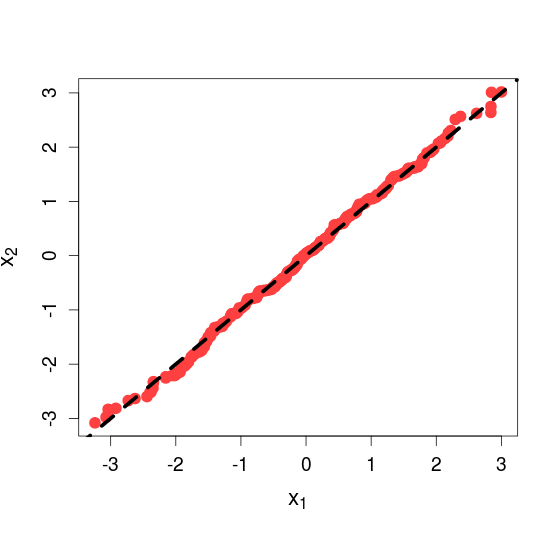}
  \includegraphics[width=0.4\textwidth,natwidth=500,natheight=500]{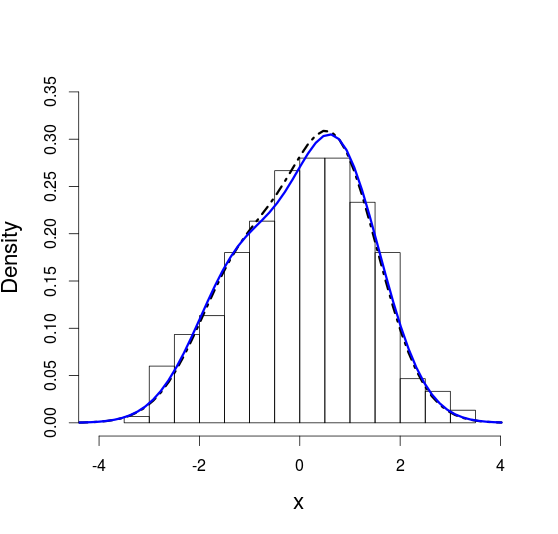}
  \caption{QQ plot, and histogram with fitted density plots}\label{Norm_ex1_fig}
   \end{center}
 \end{figure}
\end{example}
Different methods have been proposed  for determining the order of a finite mixture model, including graphical, Bayesian, penalized
likelihood maximization, and likelihood ratio hypothesis testing (\citealp{Mclachlan2000}; \citealp{Hall2005}; \citealp{Li2010};
\citealp{Maciejowska2013}).   We question though if the order is, fundamentally,  an estimable quantity: 
\begin{itemize} \item[(I)] first, mixture components may be too  close to one another to be  resolved with a given set of data, as in  Example \ref{LMM_def}. \item[(II)] Secondly,   for any fixed sample size  the mixing proportion for some components may be so
small that  contributions  from these  components may  not be observed.  
\end{itemize}For instance,
\cite{Culter1994} show, using an extensive simulation, that when the sample size is small or the components are not well
separated,  likelihood based and penalized likelihood based methods tend to overestimate or underestimate this parameter.
\cite{Donoho1988}, studies the order as a functional of a mixture density and points out that, ``near any distribution of interest,
there are empirically indistinguishable distributions (indistinguishable at a given sample size) where the functional takes on
arbitrarily large values''. He adds, ``untestable prior assumptions would be necessary'', additional to the empirical data,
for placing an upper bound. \cite{Celeux2007} mentions that this problem is weakly identifiable from data as two mixture models
with distinct number of components might not be distinguishable. 

This fundamental identification issue has immediate consequences when we are trying to define a tractable  parameter space.
In particular its dimension is problematic: the space will  have many dimensional  singularities as component points merge
or mixing distributions become singular.  Identifiability with mixtures has been well studied  of course,
see \cite{tallis1969identifiability} and \cite{lindsay1993uniqueness}. The boundaries and  singularities  in the parameter
space of a finite mixture have been looked at in \cite{leroux1992consistent}, \cite{chen1996penalized}  and \cite{Li2008}
as have the corresponding effects on the shape of the  log-likelihood function, for example see   \cite{gan1999test}.

\subsection{The Local Mixture Approach}\label{The local mixture approach} 

Examples where there is a single  set of  closely grouped components -- or the  much more general situation where $Q$ is
any small-variance  distribution -- motivated  the design of the local mixture model (LMM), \cite{Marriott2002},
\cite{Anaya-Izquierdo2007}. This is constructed around  parameters about which there is information in the data and can
be justified by a Laplace,  or Taylor, expansion argument.
\begin{definition}\label{LMM_def}
For a mean, $\mu$, parametrized density $f(x;\mu)$ belonging to the regular
exponential family, the local mixture of order $k$ is defined as 
\begin{eqnarray}\label{expression of LMM}
g_{\mu}(x; \lambda):=f(x;\mu)+\sum\nolimits_{j=1}^{k} \lambda_j f^{(j)}(x;\mu) \hspace{.5cm} \lambda \in \Lambda_{\mu}
\subset \mathbb{R}^{k} \label{lmm_4}
\end{eqnarray}
where $\lambda=(\lambda_1,\cdots,\lambda_k)$ and $f^{(j)}(x;\mu) =\frac{\partial^j f}{\partial \mu^j}(x;\mu)$. 
We denote $q_{j}(x;\mu):=\frac{f^{(j)}(x;\mu)} {f(x;\mu)}$, then for common sample space $S$, and any fixed $\mu$, 
$$\Lambda_\mu=\left\{\lambda | 1+\sum\nolimits_{j=1}^k \lambda_j \, q_{j}(x;\mu)\geq 0, \forall  x\in S\right\},$$
is a convex subspace obtained by intersection of half-spaces.  The boundary of $\Lambda_\mu$ 
corresponds to  a positivity condition on $g_\mu(x;\lambda)$.
\end{definition}

\begin{example}[{\bf \ref{Norm_ex1} revisited}] The right panel of Fig.~\ref{Norm_ex1_fig} shows the LMM fit to  the two datasets considered above. We see that the model  can successfully capture the shape of the data using only a small number of parameters about which the data is informative.
\end{example}

The local mixture approach is designed, using geometric principles, to generate an excellent inferential frame in the situation which motivated it. The `cost' associated with these properties is having to work explicitly with boundaries in the inference. We give more details of these properties and the tools associated with working with the boundaries  in \S \ref{Local and global mixture models}.  Of course the major weakness of this approach is that it says nothing when the mixing is not `local'. This paper addresses this issue by looking at finite mixtures of local mixture models. This combines the nice properties of finite mixtures, for example  the work of \cite{Lindsay1995}, while avoiding the fundamental trap of overidentifying the models as described in \S \ref{Background}.  We use this finite mixture of local mixtures to approximate  the parameter space of all mixtures. In later sections estimation methods in this very rich model class are discussed, as is  the problem of what a particular data set  can tell us about the number of components examined in  important classes of mixture models.

\section{Local and Global Mixture Models}\label{Local and global mixture models}

 Let us consider a general mixture model  of the form $\int_{\mu \in M} f(x; \mu) dQ(\mu)$ where we make the assumption that the support of $Q$, $M$, is compact. We can therefore partition $M$ as $M = \cup_{i=1}^L M_i$ where $M_i \cap M_j = \emptyset$ for $i \ne j$, and each $M_i$ is connected. Let us also select a set of `grid points', $\mu_i \in M_i$, which will be fixed and known throughout. 

The distribution $Q$ can be written as a convex combination of  distributions
 $Q = \sum_{i=1}^L \rho_i Q_i,$ where (i) $Q_i$ has support $M_i$,  and (ii) for large enough $L$ each $Q_i$ is a localising mixture in the sense required by  \cite{Anaya-Izquierdo2007}, allowing each term $\int_{\mu \in M_i} f(x; \mu) dQ_i(\mu)$ to be well approximated by a LMM. 
In the form given in Definition \ref{LMM_def}  the mean of the LMM is $\mu+\lambda_1$, so there is one degree of 
ambiguity about the parametrisation $(\mu, \lambda)$. In \cite{Anaya-Izquierdo2007} this was resolved by always setting
$\lambda_1=0$. In Definition \ref{def_FMLMM} the mean ambiguity is resolved by fixing $\mu_i$ and having $\lambda_1^i$ free.   
\begin{definition}\label{def_FMLMM}
Let  $g_{\mu_l}(x;\lambda^l)$ be  the LMM from Definition \ref{LMM_def}, and  $\lambda^l=(\lambda_1^l,\cdots,
\lambda_k^l)$. A discrete mixture of LMMs is defined by 
\begin{eqnarray}
h(x,\boldmath{\mu},\rho,\lambda)=\sum\nolimits_{l=1}^{L}\rho_l\, g_{\mu_l}(x;\lambda^l) 
\end{eqnarray}
where $\lambda=(\lambda^1,
\cdots,\lambda^L)$, $\mu=(\mu_1,\cdots,\mu_L)$ is a fixed grid of support points, $\rho=(\rho_1,\cdots,\rho_L)$ such
that $0\leq \rho_l\leq 1$ and $\sum \nolimits_{l=1}^{L}\rho_l =1$. 
\end{definition}

There are some points to consider in this definition. First, the choice of
how to select the fixed  grid points $\mu_i$, in particular how far they should  be separated, is clearly critical and 
discussed  in \S \ref{support points}. Second, throughout this paper we only  consider LMMs of order $k=4$. Increasing this degree -- while mathematically possible -- only adds a  small marginal improvement to the  local modelling performance, (\citealp{Marriott2006}). Third, whenever $f(x;\mu)$ is a proper  exponential family,  the terms $q_j(x,\mu)$'s are polynomials of degree $j$, and
the interior of the parameter space $\Lambda_{\mu_0}$ can be characterized by analysing the roots 
of a quartic polynomial. Finally, we use throughout  two illustrative examples: the normal and binomial. 
\begin{example}[Normal]\label{Normal_ex}
For the normal density function $\phi(x;\mu,1)$, with fixed variance at $\sigma^2=1$, the LMM at $\mu=\mu_0$ has the following
form,
\begin{eqnarray}
g_{\mu_0}(x;\lambda)&=&\phi(x;\mu_0,1)\{ 1+ \lambda_1 (x-\mu_0)+ \lambda_2[(x-\mu_0)^2-1]+ \lambda_3 [(x-\mu_0)^3\nonumber\\
&&-3(x-\mu_0)] +\lambda_4[(x-\mu_0)^4-6(x-\mu_0)^2+3]\} \label{Norm_ex}
\end{eqnarray}
$$\hspace{-1.5cm}\text{with}, \hspace{1cm}E(X)=\mu_0+\lambda_1, \hspace{.5cm} Var_g(X)=1+2\lambda_2-\lambda_1^2, \hspace{.5cm}
\mu_g^{(3)}=6\lambda_3+2\lambda_1^3-6\lambda_1\lambda_2$$
in which $\mu_g^{(3)}$ is the third central moment. The expression for the first moment and an argument based on Fisher
orthogonality of density derivatives (\citealp{Morris1982}) illustrate how identifiability is attained either by fixing
$\mu=\mu_0$ or $\lambda_1=0$. 
\end{example}

\begin{example}[Binomial]\label{Binomial_ex}
The LMM for a binomial distribution, with mean $\mu=\mu_0$ and number of trials $n$, has a probability mass function
of the form 
\begin{eqnarray}
g_{\mu_0}(x;n,\lambda)&=& \frac{n!\mu_0^x (n-\mu_0)^{n-x}}{x! (n-x)!n^n} \{ 1+ \lambda_1 p_1(x,\mu_0)+ \lambda_2p_2(x,\mu_0)+
\lambda_3 p_3(x,\mu_0)\nonumber\\  &&+\lambda_4p_4(x,\mu_0)\} \label{binom__ex}
\end{eqnarray}
where $p_j(x,\mu_0)$ is a polynomial with degree $j$. In this example there is extra boundary structure as $\mu$ is limited to
the compact set $[0,n]$. 
\end{example}


\begin{definition}\label{Hard Boundary}
For fixed $\mu_0$ the parameter space $\Lambda_{\mu_0}$ is a convex subset of ${\mathbb R}^4$ and its boundary, $\partial\Lambda_{\mu_0}$  is defined by the envelope of hyperplanes
$$\Pi_x:= \left\{\lambda | 1+\sum\nolimits_{j=1}^4 \lambda_j \, q_{j}(x;\mu)= 0\right\},$$ parametrized by $x\in S$, \cite{Struik1988}.
\end{definition}

\subsection{Choosing the Support Points}\label{support points}

In  Definition \ref{def_FMLMM}  the set of support points, $\{\mu_1,\cdots, \mu_L\}$, is assumed fixed and the question
remains:  how to select it?  Recall that the LMM gives a good approximation when the variance of the  mixing distribution
is small. This would imply that  we want neighbouring   support points to be close, on the other hand the more support
points the larger the value of $L$ and hence the larger the dimension of the parameter space in    Definition \ref{def_FMLMM}.

%
%

The following result follows from standard  Taylor remainder results and formalizes the above discussion.

\begin{lemma}\label{del_epsi}
Suppose $g_{\mu_0}(x;\lambda)$ is the local mixture of the family of densities $f(x;\mu)$ and $Q$ is a distribution.  For any $\delta >0$ there
exist an interval $I=[\mu_0-\epsilon_1(\delta) ,\mu_0+\epsilon_2(\delta)]$ 
such that 
$$\left|\int_I f(x;\mu)\,dQ-g_{\mu_0}(x;\lambda)\right|<\delta,$$ for all $x$.
\end{lemma}

\begin{example}[{\bf \ref{Normal_ex} revisited}] 
By Taylor's theorem we have 
$f(x;\mu)-g_{\mu_0}(x;\lambda(\mu))= \frac{(\mu-\mu_0)^5}{5!}f^{(5)}(x;m)$ where $m$ is a value between $\mu$ and $\mu_0$. 
For the normal family with standard deviation $\sigma$ we have 
$$f^{(5)}(y,m)=\left( y^5-10\frac{y^3}{\sigma^2}+15\frac{y}{\sigma^4}\right)\frac{e^{-\frac{y^2}{2\sigma^2}}}{\sqrt{2\pi}\sigma},$$
where $y=\frac{(x-m)}{\sigma^2}$. This function is obviously bounded, by $M$ say,  for all $y\in \mathbb{R}$, and the bound, which only depends on $\sigma$,
can be numerically obtained. Letting  $\epsilon=\max\{\epsilon_1,\epsilon_2\}$ gives,
\begin{eqnarray}
\left|\int_I f(x;\mu)\,dQ-\int_I g_{\mu_0}(x;\lambda(\mu))\,dQ\right|    &\leq& \int_I |f(x;\mu)-g_{\mu_0}(x;\lambda)|\,dQ\nonumber\\
&\leq&  (\epsilon_1+\epsilon_2)\,\frac{\epsilon^5}{5!}\,M
\end{eqnarray} The  result follows since  we can write $\int_I g_{\mu_0}(x;\lambda(\mu))\,dQ$ as a LMM with
$\lambda_i := \int \lambda_i(\mu) dQ(\mu)$.
\end{example}

\begin{example}[{\bf \ref{Binomial_ex} revisited}] 
For the binomial family, with probability function $p(x;n,\mu)$, again we want to bound the error by  $\frac{(\mu-\mu_0)^5}{5!}M$, say. We have 
$$p^{(5)}(x;n,m)=p(x;n,m)\, q_5(x;n,m)$$
where $q_5(x;n,m)$ is a polynomial of degree $5$ of both $x$ and $m$, which can be written as 
$$q_5(x;n,m)=\frac{1}{(n-m)^5}\sum_{j=0}^{5}\gamma(j){5 \choose j}(\frac{n}{m})^j (-1)^{5-j}$$
with $\gamma(j)=j!\,(5-j)! {x \choose j} {n-j \choose n-5}$. It can be shown that uniformly for all $x=0,1,\cdots,n$,
$p(x;n,m)\leq p(x^{\star};n,m)$, where $x^{\star}=\lfloor\frac{m(n+1)}{n}\rfloor$, and 
$$  L(n,m) < q_5(x;n,m) < U(n,m)$$
where, for all $m\in[0,n]$,
\begin{eqnarray}
L(n,m)&=&-\frac{\gamma(0)}{(n-m)^5 m^4}(5 n^4+10 n^2 m^2+m^4)\nonumber\\
U(n,m)&=&\frac{\gamma(0)}{(n-m)^5 m^5}(n^5+10 n^3 m^2+5n m^4-m^5)
\end{eqnarray}
Moreover, it can be shown that 
$$
\left\{ \begin{array}{ll}
         U(n,m)>|L(n,m)| & \mbox{if \,\,\,\,$0 \leq m\leq \frac{n}{2}$};\\
        U(n,m)<|L(n,m)| & \mbox{if \,\,\,\,$\frac{n}{2}< m\leq n$}.\end{array}\right.
$$
Therefore,
$$M=\max_{m \in I}q_5(x^{\star};n,m)|L(n,m)|\,\,\,\,\,\, or\,\,\,\,\,\, M=\max_{m \in I}q_5(x^{\star};n,m)U(n,m)$$
which depends on $\mu_0$, $\epsilon_1$ and $\epsilon_2$.$\square$
\end{example}

%


\subsection{Estimation Methods}

Estimation with a LMM is, in general,  straightforward. The  parameter space has nice  convexity properties and  the likelihood is  log-concave,
see \cite{Anaya-Izquierdo2007}. In \cite{Marriott2002} Markov Chain Monte Carlo (MCMC) methods are used since boundaries
in the parameter  space can easily be accommodated by a simple rejection step whenever a parameter value is proposed that
lies outside the boundary. Alternatively direct log-likelihood maximization can be done exploiting the convexity of the
parameter space and the concavity of the objective function (\citealp{Maroufy}). 

Adopting these ideas  to finite mixtures of LMMs, we  can also easily use MCMC methods. However,  here  we define  a new  form
of Expectation-Maximization (EM) algorithm, described in the Appendix, and applied in  Example \ref{Acidity_new}.  In this
example we look at mixtures of normals, $\phi(x; \mu, \sigma_0^2)$,  where grid-points for $\mu$ are selected as discussed in
\S \ref{support points}. To understand the selection of $\sigma_0^2$ by the modeller  we   return to  point (II) of Section
\ref{Background}. This makes the case that we can only estimate clusters, and indeed features of such clusters, if there is
the associated information in the data.   One consequence of that is the well-known phenomenon that infinite likelihoods
are attainable in the case where  only a single  observation has been associated with a normal cluster and the estimated variance
is zero.  In our approach we take issue (II)  seriously and only put in a LMM component when there is enough data to support its
inference.  In particular we note that the variance of such a component is  $\sigma_0^2+ 2\lambda_2-\lambda_1^2$, which will be bounded below,
and vary from cluster to cluster.  Hence the data can estimate the variance of each cluster as long as it is above our, modeller
selected, threshold.

\begin{example}[Acidity data]\label{Acidity_new}
The data includes acidity index measured in 155 lakes in north-central Wisconsin which is analyzed in \cite{Mclachlan2000}
and the references therein. Using likelihood ratio hypothesis testing,
the bootstrap estimated p-value supports two or three components at the $5\%$ or $10\%$ level of significance,
respectively. However, based on a Bayesian method all the values between two and six are   equally supported, \citealp{Richardson1997}.

Here we select the grid-points  $\mu^{(0)}=(3.6,4.2,4.8,5.4,6,6.6,7)$, set  $\sigma_l=0.5$ and $\gamma=0.15$, so that at least $20$ observation is
assigned to each cluster. The algorithm returns a two-component discrete mixture of LMMs with $\hat{\rho}=(0.676,0.324)$,
$\mu=(4.2,6.6)$, Figure \ref{Acidity_EM} (left panel). The middle panel shows that if we give a set of slightly different
initial grid points, $\mu^{(0)}_6=6.4$ instead of $6.6$, the algorithm returns the same order for the mixture, with $m=(4.2,6.4)$
and $\hat{\rho}=(0.651,0.349)$, (middle panel). In addition, if we let $\sigma_l$'s to take different values, $\sigma_6=0.6$, 
we get the same order with a slightly different fit (right panel).

\begin{figure}[!h]  
 \includegraphics[width=0.32\textwidth,natwidth=410,natheight=442]{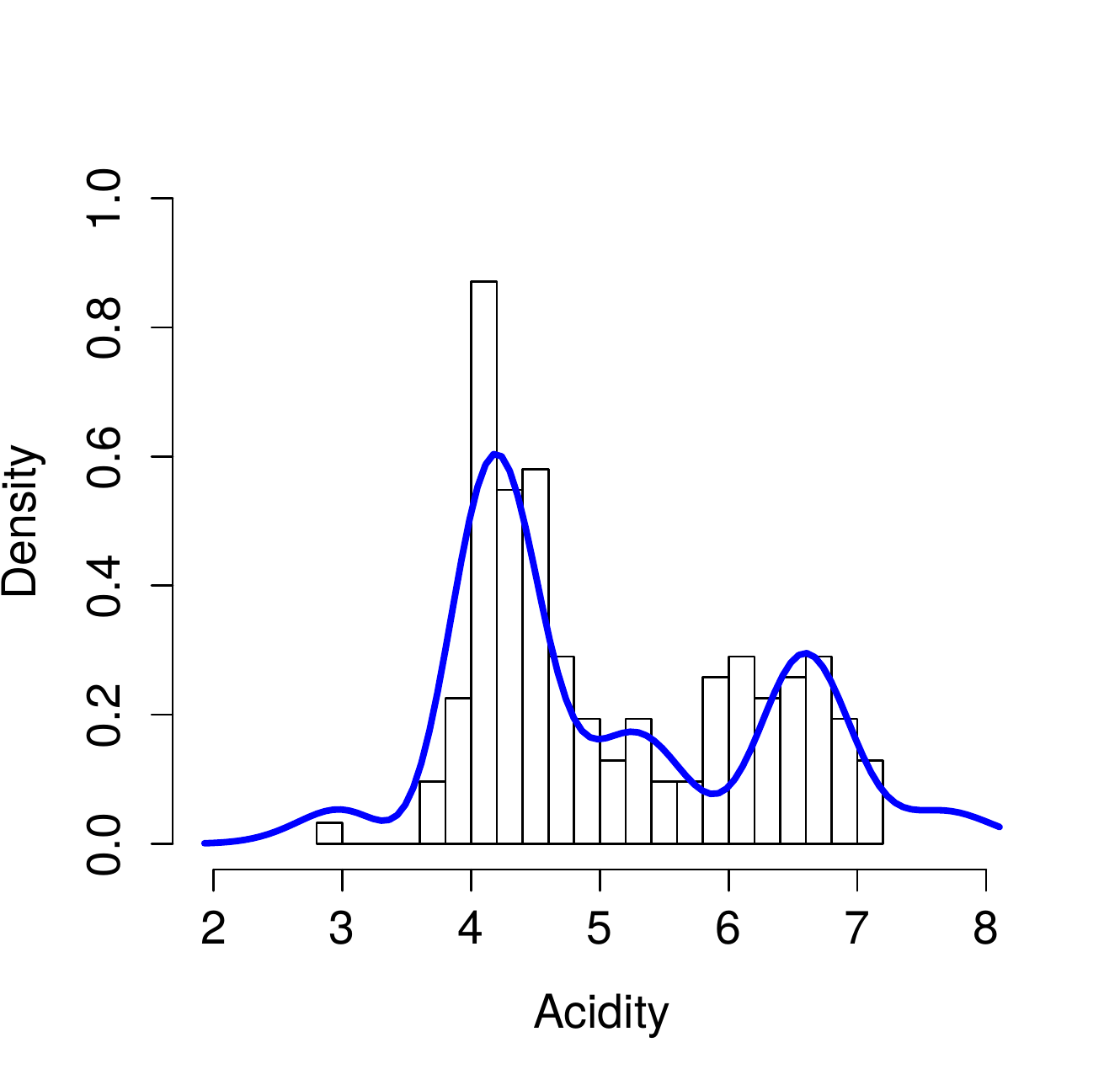}
 \includegraphics[width=0.32\textwidth,natwidth=410,natheight=442]{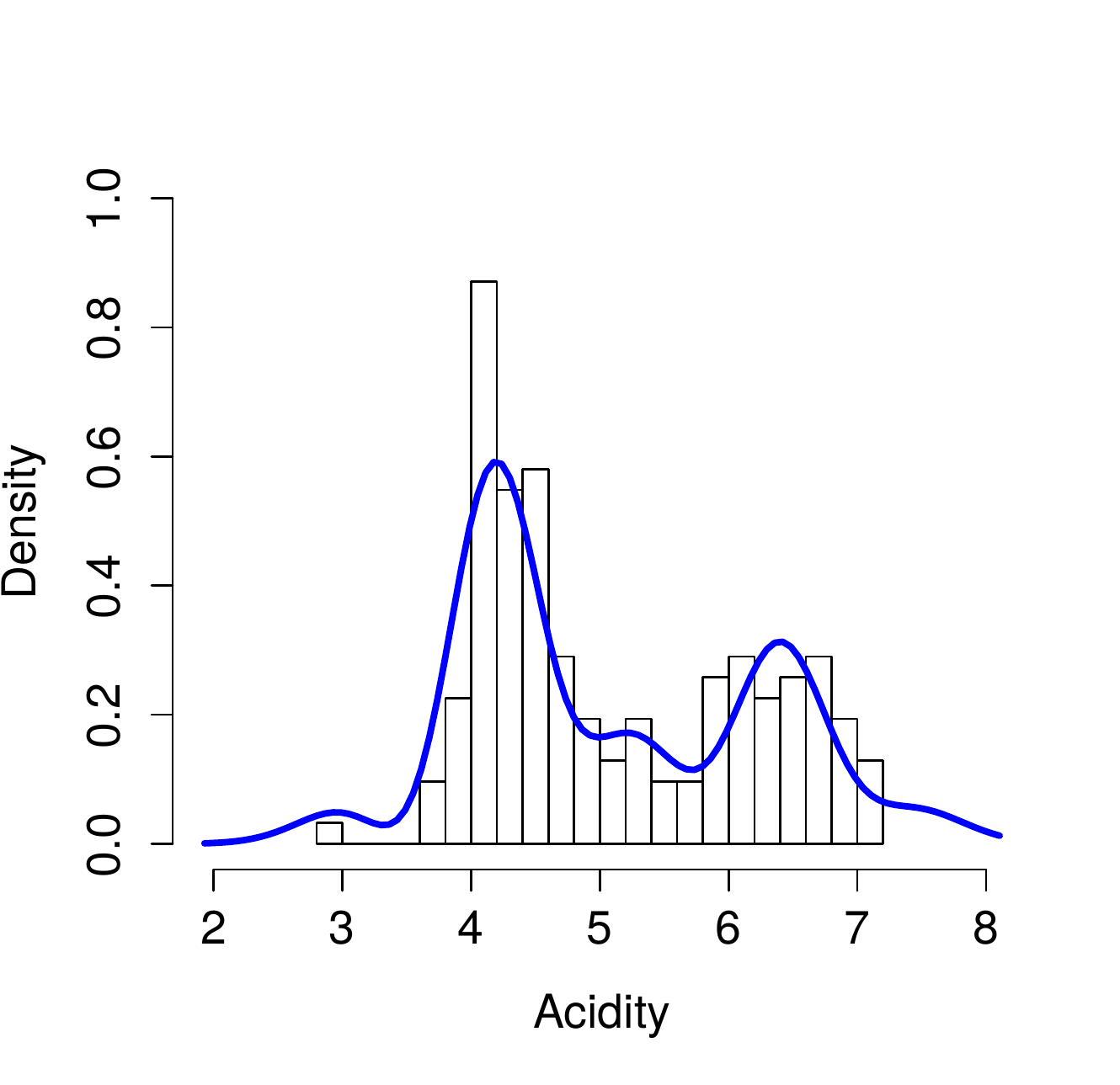}
 \includegraphics[width=0.32\textwidth,natwidth=410,natheight=442]{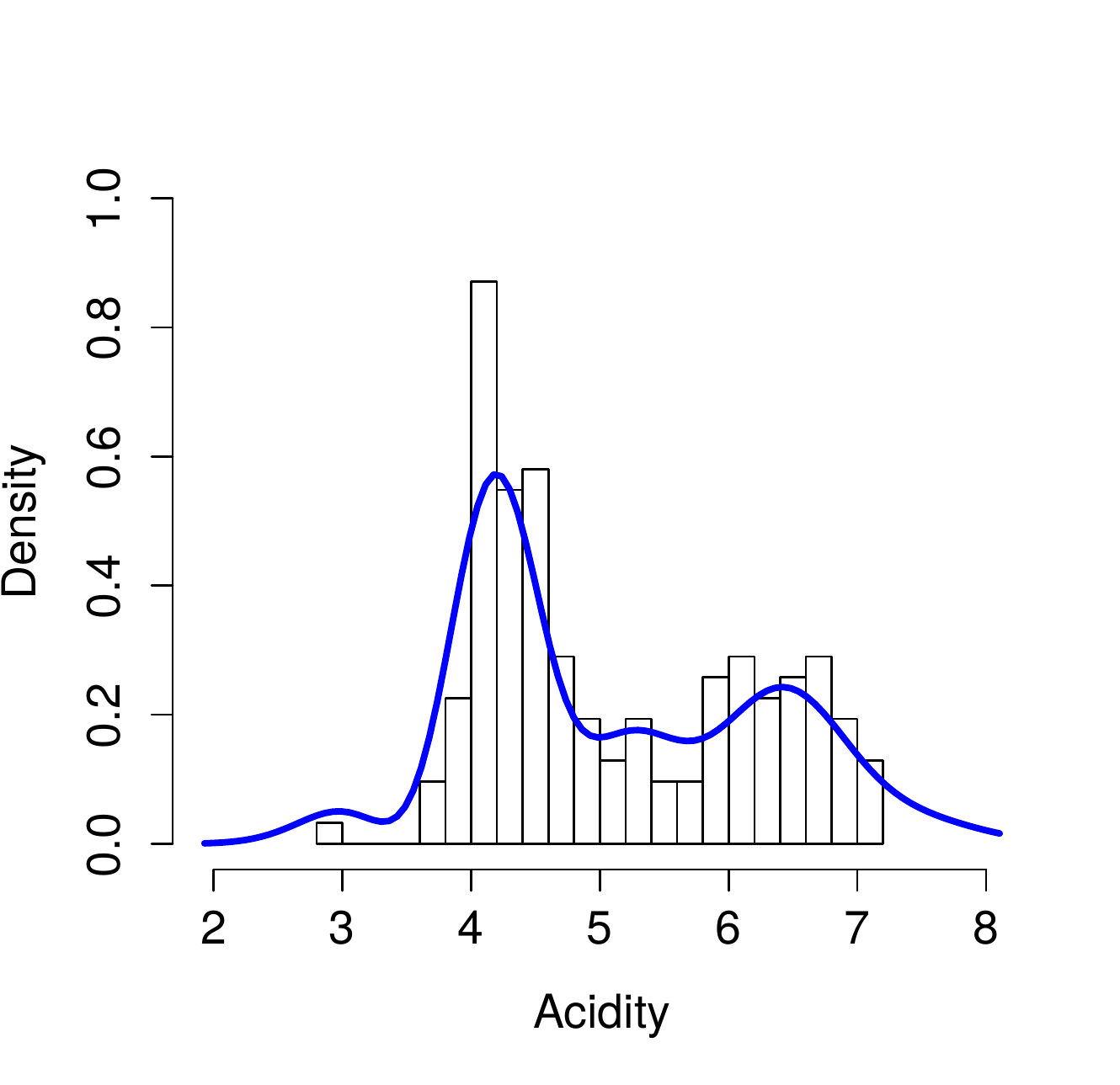}
 \caption{Discrete mixture of LMMs for Acidity data.}\label{Acidity_EM}
\end{figure} 
Further analysis of the data with different values of $\gamma$, shows how the final results of the algorithm
depend on  $\gamma$, see Table \ref{gamma_table}.
\begin{table}[h]
\begin{center}
    \begin{tabular}{ | p{4.5cm} | p{3cm} | p{4cm} | l |}
    \hline
    $\gamma$ & $\mu$ & $\hat{\rho}$ & Order \\ \hline\hline
    $0.13,0.14,0.15,0.16,0.17$ & $(4.2,6.6)$ & $(0.67,0.33)$ & 2 \\ \hline
    $0.1,0.11,0.12$ & $(4.2,4,8,6.6)$ & $(0.57,0.13,0.3)$ & 3 \\ \hline
    $0.07,0.08,0.09$ & $(4.2,6,6.6)$ & $(0.63,0.18,0.19)$ & 3 \\ \hline
    $0.06$ & $(4.2,4.8,6,6.6)$ & $(0.57,0.08,0.16,0.19)$ & 4\\
    \hline
    \end{tabular} \caption{Further analysis for different values of $\gamma$}
    \label{gamma_table}
\end{center}
 
\end{table}
\end{example}

\begin{figure}[!h]  
 \includegraphics[width=0.32\textwidth,natwidth=410,natheight=442]{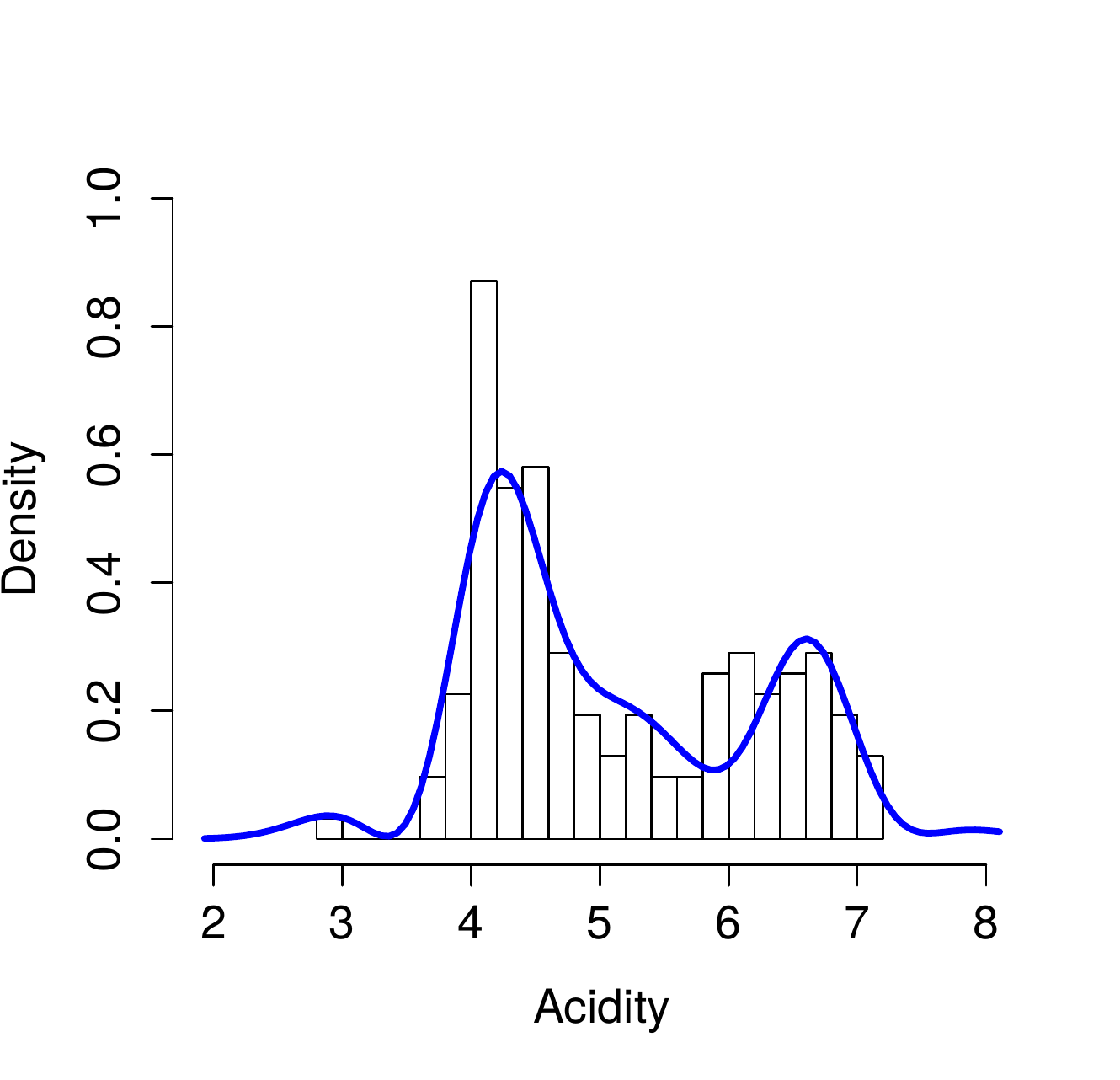}
 \includegraphics[width=0.32\textwidth,natwidth=410,natheight=442]{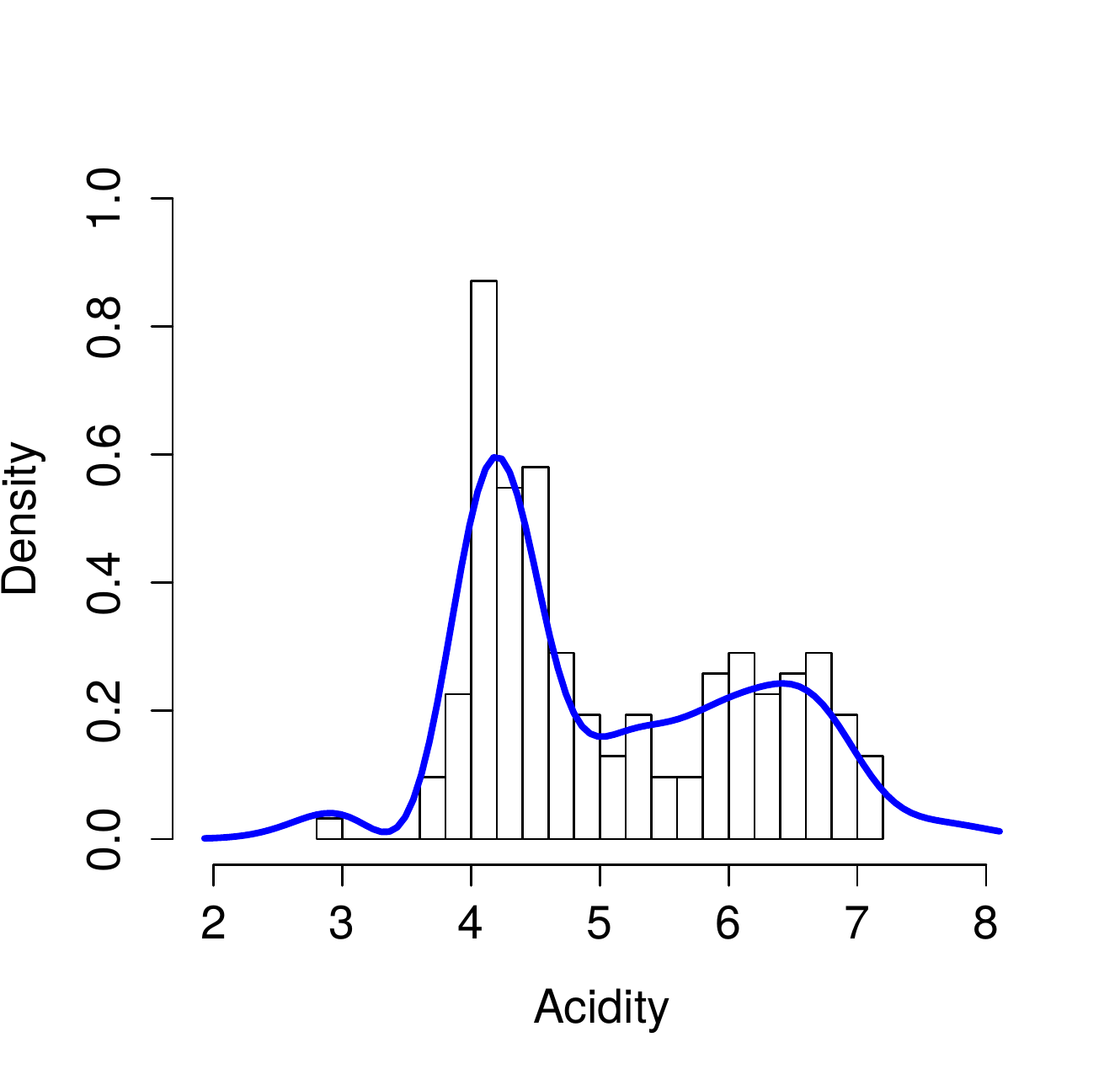}
 \includegraphics[width=0.32\textwidth,natwidth=410,natheight=442]{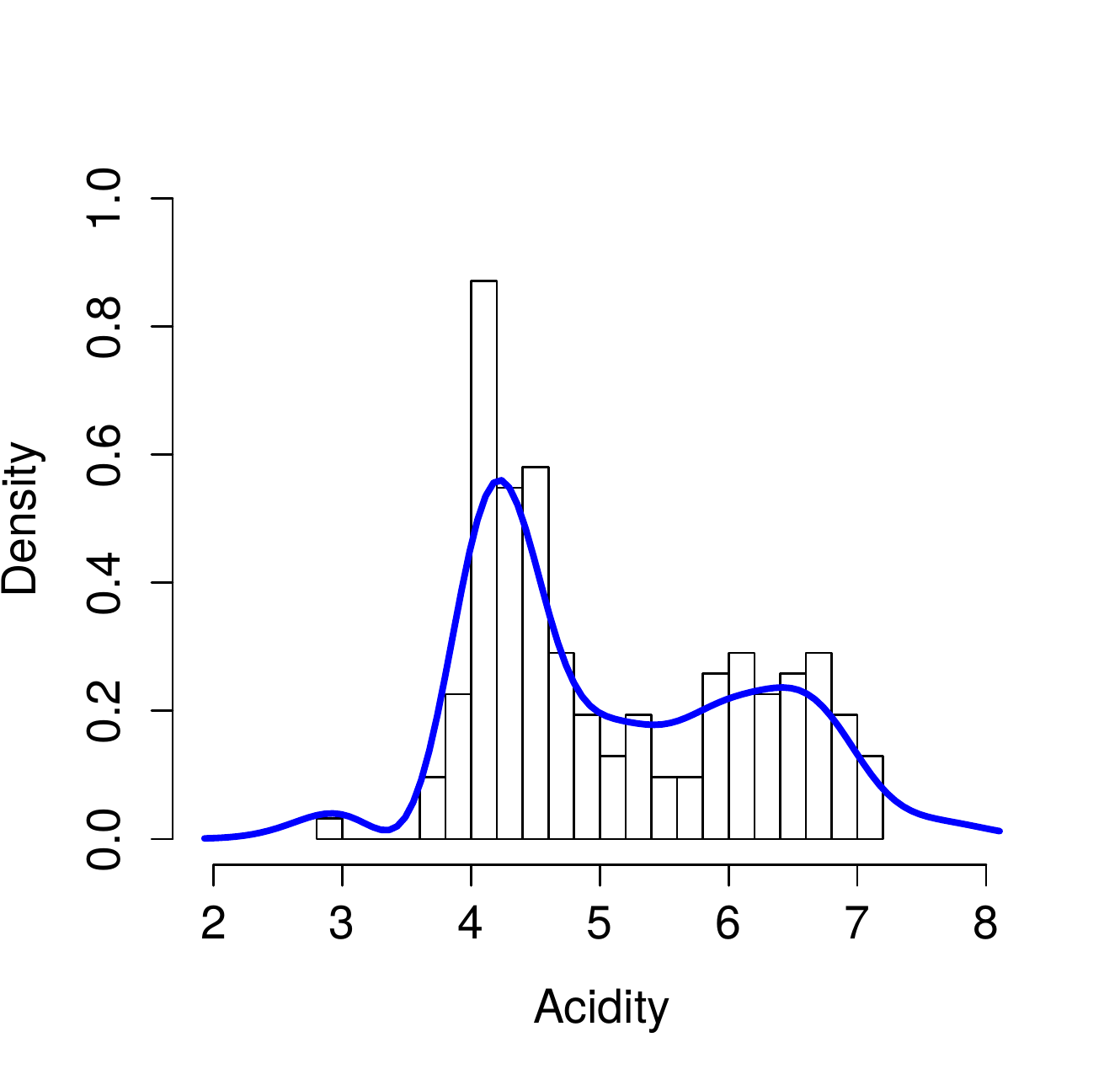}
 \caption{left to right: three and four components fit corresponding to the last three rows in
 Table \ref{gamma_table}}\label{Acidity_Extra}
\end{figure}

\section{Discussion}

While finite mixtures of exponential families are very flexible they suffer from identification  problems when support points cluster.
This means estimating the order is a very hard problem with a fixed set of data.   This paper takes a new approach to this problem.
We use a local mixture model   to directly model each  cluster in a very flexible way. This  results in a finite mixture of  LMMs.  We  propose counting  these,
now well-defined,  components  as the `order' -- which will now be statistically meaningful.  In each of the component LMMs  all  the
parameters are estimable with efficient algorithms where we have applied a principle that we do not considered models which are unestimable from the data at hand.

\bibliographystyle{abbrvnat}
\bibliography{jabref.bib}

\begin{thebibliography}{24}
\providecommand{\natexlab}[1]{#1}
\providecommand{\url}[1]{\texttt{#1}}
\expandafter\ifx\csname urlstyle\endcsname\relax
  \providecommand{\doi}[1]{doi: #1}\else
  \providecommand{\doi}{doi: \begingroup \urlstyle{rm}\Url}\fi

\bibitem[Anaya-Izquierdo and Marriott(2007)]{Anaya-Izquierdo2007}
K.~Anaya-Izquierdo and P.~Marriott.
\newblock Local mixture models of exponential families.
\newblock \emph{Bernoulli}, 13:\penalty0 623--640, 2007.

\bibitem[Celeux(2007)]{Celeux2007}
G.~Celeux.
\newblock Mixture models for classification.
\newblock \emph{Advances in Data Analysis}, Springer Berlin
  Heidellberg:\penalty0 3--14, 2007.

\bibitem[Chen and Kalbfleisch(1996)]{chen1996penalized}
J.~Chen and J.~Kalbfleisch.
\newblock Penalized minimum-distance estimates in finite mixture models.
\newblock \emph{Canadian Journal of Statistics}, 24\penalty0 (2):\penalty0
  167--175, 1996.

\bibitem[Culter and Windham(1994)]{Culter1994}
A.~Culter and Windham.
\newblock Information-based validity functionals for mixture analysis.
\newblock \emph{In Proceedings of the First US/Japan Conference on the
  frontires of Statistical Modeling in Informational Approach}, Amsterdam:
  Kluwer:\penalty0 149--170, 1994.

\bibitem[Donoho(1988)]{Donoho1988}
D.~L. Donoho.
\newblock One-sided inference about functionals of a density.
\newblock \emph{Annals of statistics}, 16:\penalty0 1390--1420, 1988.

\bibitem[Everitt(1996)]{everitt1996introduction}
B.~S. Everitt.
\newblock An introduction to finite mixture distributions.
\newblock \emph{Statistical Methods in Medical Research}, 5\penalty0
  (2):\penalty0 107--127, 1996.

\bibitem[Gan and Jiang(1999)]{gan1999test}
L.~Gan and J.~Jiang.
\newblock A test for global maximum.
\newblock \emph{Journal of the American Statistical Association}, 94\penalty0
  (447):\penalty0 847--854, 1999.

\bibitem[Hall and Stewart(2005)]{Hall2005}
P.~Hall and M.~Stewart.
\newblock Theoretical analysis of power in a two-componet normal mixture model.
\newblock \emph{Journal of Statistical Planning and Inference}, 134:\penalty0
  158--179, 2005.

\bibitem[Leroux et~al.(1992)]{leroux1992consistent}
B.~G. Leroux et~al.
\newblock Consistent estimation of a mixing distribution.
\newblock \emph{The Annals of Statistics}, 20\penalty0 (3):\penalty0
  1350--1360, 1992.

\bibitem[Li and Chen(2010)]{Li2010}
P.~Li and J.~Chen.
\newblock Testing the order of a finite mixture.
\newblock \emph{Journal of the American Statistical Association},
  105:491:\penalty0 1084--1092, 2010.

\bibitem[Li et~al.(2008)Li, Chen, and Marriott]{Li2008}
P.~Li, J.~Chen, and P.~Marriott.
\newblock Non-finite fisher information and homogeneity: an em approach.
\newblock \emph{Biometrika}, pages 1--16, 2008.

\bibitem[Lindsay(1995)]{Lindsay1995}
B.~G. Lindsay.
\newblock \emph{Mixture Models: Theory, Geometry and Applications}.
\newblock Inst of Mathematical Statistics, 1995.

\bibitem[Lindsay and Roeder(1993)]{lindsay1993uniqueness}
B.~G. Lindsay and K.~Roeder.
\newblock Uniqueness of estimation and identifiability in mixture models.
\newblock \emph{Canadian Journal of Statistics}, 21\penalty0 (2):\penalty0
  139--147, 1993.

\bibitem[Maciejowska(2013)]{Maciejowska2013}
K.~Maciejowska.
\newblock Assessing the number of componentsi a normal mixture: an alternative
  appraoch.
\newblock \emph{University Library of Munich}, \penalty0 (No. 50303), 2013.

\bibitem[Maroufy and Marriott(2015)]{Maroufy}
V.~Maroufy and P.~Marriott.
\newblock Generalizing the frailty assumptions in survival analysis.
\newblock \emph{arXiv:1510.02425}, 2015.

\bibitem[Marriott(2002)]{Marriott2002}
P.~Marriott.
\newblock On the local geometry of mixture models.
\newblock \emph{Biometrika}, 89:\penalty0 77--93, 2002.

\bibitem[Marriott(2006)]{Marriott2006}
P.~Marriott.
\newblock Extending local mixture models.
\newblock \emph{AISM}, 59:\penalty0 95--110, 2006.

\bibitem[Mclachlan and Peel(2000)]{Mclachlan2000}
G.~Mclachlan and D.~Peel.
\newblock \emph{Finite Mixture Models}.
\newblock John Wiley and sons, 2000.

\bibitem[Morris(1982)]{Morris1982}
C.~Morris.
\newblock Natural exponential families with quadratic variance functions.
\newblock \emph{The Annals of Statistics}, 10\penalty0 (1):\penalty0 65--80,
  1982.

\bibitem[Richardson and Green(1997)]{Richardson1997}
S.~Richardson and P.~J. Green.
\newblock On bayesian analysis of mixtures with an unknown number of components
  (with discuassion).
\newblock \emph{Journal of the Royal Statistical Society B}, 59:\penalty0
  731--792, 1997.

\bibitem[Schlattmann(2009)]{Schlattmann2009}
P.~Schlattmann.
\newblock \emph{Medical Applications of Finite Mixture models}.
\newblock Springer, 2009.

\bibitem[Shun and McCullagh(1995)]{shun1995laplace}
Z.~Shun and P.~McCullagh.
\newblock Laplace approximation of high dimensional integrals.
\newblock \emph{Journal of the Royal Statistical Society. Series B
  (Methodological)}, pages 749--760, 1995.

\bibitem[Struik(1988)]{Struik1988}
D.~J. Struik.
\newblock \emph{Lectures on Classical Differential Geometry}.
\newblock Dover Publications, 1988.

\bibitem[Tallis(1969)]{tallis1969identifiability}
G.~Tallis.
\newblock The identifiability of mixtures of distributions.
\newblock \emph{Journal of Applied Probability}, 6\penalty0 (2):\penalty0
  389--398, 1969.

\end{thebibliography}

\appendix

\section*{Appendix}\label{Appendix}

Starting from  initially selected grid points $\mu^{(0)}=(\mu_1^{(0)},\cdots,\mu_L^{(0)})$,  proportions  
$\rho^{(0)}=(\rho_1^{(0)},\cdots,\rho_L^{(0)})$  and local mixture parameters $\lambda^{(0)}=(\lambda^{1,(0)},\cdots,\lambda^{L,(0)})$. Suppose we have 
$\mu^{(r)}$ and $\rho^{(r)}$ and $\lambda^{(r)}$ at step $r$, where $L_r\leq L$. For obtaining the estimates at step $r+1$
run the following steps.

\begin{enumerate}
 \item Calculate $\rho^{(r+1)}=\frac{n_l}{n}$, where $n_l= \sum\nolimits_{i=1}^{n}w^{(r+1)}_{il}$ and 
 $$w^{(r+1)}_{il}=\frac{\rho_l^{(r)} g_{\mu_l}(x_i,\lambda^{l,(r)})}{\sum\nolimits_{l=1}^{L_r}
 \rho_l^{(r)} g_{\mu_l}(x_i,\lambda^{l,(r)})}, \hspace{1cm} x=1,\cdots,n;\,\, l=1,\cdots,L_r$$

 \item Choose a positive value $0 < \gamma < 1$, and check if there is any $l$ such that $\rho_l^{(r+1)} < \gamma$.
 \begin{enumerate}
  \item If yes: exclude the components corresponding to $\rho_l^{(r+1)} < \gamma$, update $L_r\rightarrow L_{r+1}$
  and go back to step $1$.
 \item If no: go to step $3$. 
 \end{enumerate}

 \item Classify the data set into $x^1,\cdots x^{L_{r+1}}$ by assigning each $x_i$ to only one mixture component.
 For each $l=1,\cdots,L_{r+1}$, update $\lambda^{l,(r)}$ by
 $$\lambda^{l,(r+1)}= \operatorname*{arg\,max}_{\lambda \in \Lambda_{\mu_l}} l_{\mu_l}(x^l,\lambda),$$
 where $ l_{\mu_l}(x^l,\cdot)$ is the log-likelihood function for the component $l$. 
\end{enumerate}

\begin{remark}
 Step $2$ restricts the number of required components for fitting a data set in a way that there is enough information
 necessary for running inference on each local mixture component. Its value has an influence on the final result of the
 algorithm in a similar way that an initial value affects the convergence of a general EM algorithm (Table \ref{gamma_table}). 
\end{remark}

\end{document}